\documentclass[twocolumn,showpacs,preprintnumbers,amsmath,amssymb,APSl,prd,nofootinbib,superscriptaddress]{revtex4-1}

\usepackage{dcolumn}
\usepackage{bm}
\usepackage{ifpdf}
\usepackage{hyperref}
\usepackage{bm}
\usepackage{xcolor,color,graphicx,graphics}
\usepackage[spanish,english]{babel}
\usepackage[latin1]{inputenc}
\usepackage[OT1]{fontenc}
\usepackage{latexsym,amssymb,amsmath,amsfonts}
\usepackage{makeidx}
\usepackage{epsfig,subfigure}
\usepackage{natbib}
\usepackage{epstopdf}
\usepackage{mathrsfs}
\usepackage{hyperref}
\hypersetup{colorlinks=true, linkcolor=blue, citecolor=green}
\usepackage{enumerate}

\newcommand{\be}{\begin{equation}}
\newcommand{\en}{\end{equation}}
\newcommand{\bea}{\begin{eqnarray}}
\newcommand{\ena}{\end{eqnarray}}
\newcommand{\bes}{\begin{subequations}}
\newcommand{\ees}{\end{subequations}}

\begin{document}

\title{Nonsingular black holes, wormholes, and de Sitter cores from anisotropic fluids}

\author{C. Menchon}
\affiliation{Departamento de F\'{i}sica Te\'{o}rica and IFIC, Centro Mixto Universidad de Valencia - CSIC.
Universidad de Valencia, Burjassot-46100, Valencia, Spain}
\author{Gonzalo J. Olmo} \email{gonzalo.olmo@uv.es}
\affiliation{Departamento de F\'{i}sica Te\'{o}rica and IFIC, Centro Mixto Universidad de Valencia - CSIC.
Universidad de Valencia, Burjassot-46100, Valencia, Spain}
\affiliation{Departamento de F\'isica, Universidade Federal da
Para\'\i ba, 58051-900 Jo\~ao Pessoa, Para\'\i ba, Brazil}
\author{D. Rubiera-Garcia} \email{drgarcia@fc.ul.pt}
\affiliation{Instituto de Astrof\'{\i}sica e Ci\^{e}ncias do Espa\c{c}o, Faculdade de
Ci\^encias da Universidade de Lisboa, Edif\'{\i}cio C8, Campo Grande,
P-1749-016 Lisbon, Portugal}

\date{\today}
\begin{abstract}
We study Born-Infeld gravity coupled to an anisotropic fluid in a static, spherically symmetric background. The free function characterizing the fluid is selected on the following grounds: i) recovery of the Reissner-Nordstr\"om solution of GR at large distances, ii) fulfillment of classical energy conditions and iii) inclusion of models of nonlinear electrodynamics as particular examples. Four branches of solutions are obtained, depending on the signs of two parameters on the gravity and matter sectors. On each branch, we discuss in detail the modifications on the innermost region of the corresponding solutions, which provides a plethora of configurations, including nonsingular black holes and naked objects, wormholes and de Sitter cores. The regular character of these configurations is discussed according to the completeness of geodesics and the behaviour of curvature scalars.
\end{abstract}


\maketitle

\section{Introduction}

With the birth of gravitational wave astronomy following the discovery made by LIGO \cite{LIGO}, and interpreted as the merging of two astrophysical-size black holes, the theoretical and numerical understanding of black holes has acquired a renewed interested. Indeed, many proposals have arisen, where the Kerr solution of GR is replaced by more or less exotic compact objects, so as to explore potential observational signatures able to discriminate one from another \cite{Compact}. At the same time, the field offers an excellent opportunity to put to experimental test the many modifications of GR proposed in the literature, such as $f(R)$ \cite{fR}, $f(T)$ on its various formulations \cite{fT}, $f(R,T)$ \cite{fRT}, Gauss-Bonnet \cite{GB}, hybrid theories \cite{Capozziello:2015lza}, and many others \cite{MG}. See Berti et.al. \cite{Berti} for the current observational status and experimental bounds of such proposals.

Besides their interest for gravitational waves, many of such compact objects are likely to have consequences for the issue with spacetime singularities. According to the theorems on singularities developed by Penrose \cite{Penrose}, Hawking \cite{Hawking}, Carter \cite{Carter} and others (see \cite{Senovilla2015} for a pedagogical discussion), based on physically reasonable assumptions upon the causal and geometrical structure of spacetime, and which make use of the concept of \emph{geodesic completeness} (i.e., whether null and timelike geodesics can be extended to arbitrary large values of the affine parameter or not), the development of a singularity during the last stages of gravitational collapse is unavoidable within GR. To overcome this result, the literature has split into two main schools. In the first of them, one sticks to GR and tries to remove such singularities, usually paying the price of violating the energy conditions (see, however, \cite{Senovilla1990}), and then goes on to minimize it by suitably choosing the geometry, like what is usually done in thin-shell wormholes \cite{Garcia:2011aa}. In the second, one extends the GR action in looking for mechanisms able to produce a bounce during the gravitational collapse \cite{Bounce}. A natural consequence of many such mechanisms is the fact that a bound on curvature scalars arises which, consequently, has also triggered a large literature in building solutions with finite curvature scalars \cite{Bound}.

In this work we shall follow the second path and focus on a class of extensions of General Relativity inspired by the nonlinear electrodynamics of Born-Infeld \cite{BI} and termed Born-Infeld theories of gravity (see \cite{BHOR} for a recent review). On its most conventional and widely employed version, dubbed Eddington-inspired Born-Infeld (EiBI) gravity, originally introduced by  Ba\~nados and Ferreira \cite{BF} and afterwards studied by different authors in astrophysics, black hole physics and cosmology \cite{BIapp}, (null and time-like) geodesically complete spacetimes sourced by standard electromagnetic (Maxwell) fields can be found \cite{ORS16b}. Such solutions replace the point-like singularity of the Reissner-Nordstr\"om solution of GR by a wormhole structure \cite{ORS14}, which provides the mechanism for the natural extension of the geodesics without incurring in violations of energy conditions. Moreover, despite the generic presence of curvature divergences at the wormhole throat, extended objects may cross this region without experiencing destructive effects \cite{ors16}, while the problem of scattering of scalar waves off the wormhole turns out to be well posed \cite{ORS16a}. These good news seem to be tightly linked to the metric-affine (or Palatini) formulation of EiBI gravity, where metric and connection are regarded as independent entities \cite{olmo11}. Indeed, it has been shown that Palatini theories of gravity generically yield second-order equations that in vacuum reduce to the GR ones \cite{Afonso:2017bxr, Ferraris:1992dx}, this way avoiding the generic presence of ghost-like instabilities of the metric formulation of modified gravity.

Exploring further the structure of these geometries, in this work we shall refine the matter description and model it using an anisotropic fluid (i.e. having different radial and tangential pressures). Though the reliability of the isotropy in the fluid description has been experimentally verified in many contexts, there are physical arguments suggesting the appearance of anisotropies both at high and low energy densities, in particular, in realistic models of compact objects (see \cite{HS1997} for a review). These fluids have been recently employed in the study of realistic magnetized accretion disks around Kerr black holes \cite{Font2017} (see \cite{anisotropicGR} for further studies on black holes/wormholes from anisotropic fluids). In the context of EiBI gravity, recently Shaikh \cite{Shaikh2015} (see also \cite{Harko2015} for a slightly different approach to this issue) considered a simplified model for an anisotropic fluid, finding the existence of both wormholes and non-singular solutions with similar properties as those supported by electromagnetic fields above.

In this work we shall go beyond those results, and consider an anisotropic fluid with an ansatz mainly motivated by three reasons: i) recovery of the Reissner-Nordstr\"om solution of GR for far distances, ii) fulfillment of classical energy conditions and iii) inclusion of nonlinear electrodynamics as particular cases of that fluid. Our analysis will be split into four branches, according to the signs of two parameters on the gravitational and matter sectors, respectively, and we will characterize in detail the innermost region of each of the corresponding configurations on each branch. In particular, we shall devote special attention to the geodesic structure of those internal regions, and compare it to the behaviour of curvature scalars there. This analysis will reveal the existence of different kinds of objects, including wormhole structures, non-singular solutions (both cloaked with horizons and naked), and de Sitter cores.

The paper is organized as follows: in Sec. \ref{sec:II} we shall specify the gravitational and matter sectors and cast the field equations in suitable form, which are subsequently solved in Sec. \ref{sec:III}. In Sec. \ref{sec:geo} we recall the main elements of geodesic behaviour in Palatini theories of gravity and particularize them to the present case. A detailed analysis of the geometric and geodesic features of the four branches of solutions is carried out in Sec. \ref{sec:V}, and we conclude in Sec. \ref{sec:VI} with a summary of the results obtained and some perspectives for future research.

\section{Theory and setup} \label{sec:II}

\subsection{Gravity sector}

The action defining Eddington-inspired Born-Infeld gravity is given by \cite{BHOR}

\begin{equation} \label{eq:BIaction}
\mathcal{S}_{EiBI}=\frac{1}{\kappa^2 \epsilon} \int d^4x \left( \sqrt{ \vert g_{\mu\nu}+\epsilon R_{\mu\nu}(\Gamma) \vert } - \lambda \sqrt{\vert g_{\mu\nu} \vert} \right) \ ,
\end{equation}
with the following definitions and conventions: $\kappa^2=8\pi G/c^4$ is Newton's constant, vertical bars denote a determinant, $\epsilon$ is EiBI parameter with dimensions of length squared, $g_{\mu\nu}$ is the spacetime metric, which is independent of the affine connection $\Gamma \equiv \Gamma_{\mu\nu}^{\lambda}$ (Palatini approach); the (symmetrized) Ricci tensor $R_{\mu\nu}(\Gamma)$ is entirely built out of the affine connection as $R_{\mu\nu}(\Gamma) \equiv {R^\alpha}_{\mu\alpha\nu}(\Gamma)$, and $\lambda$ is a parameter related to the effective cosmological constant $\Lambda_{eff}=\frac{\lambda -1}{\epsilon}$, which follows from a series expansion in terms of $\epsilon \ll 1$ of the action (\ref{eq:BIaction}) as:
\begin{eqnarray} \label{eq:gsbi}
\mathcal{S}_{EiBI} (\epsilon \ll \kappa^2)&=& \int d^4x \sqrt{-g}\left( \frac{R}{2\kappa^2}-2\Lambda_{eff} \right) \\
+ &\epsilon& \int d^4x \sqrt{-g} \left( \frac{R^2}{2} - R_{\mu\nu}R^{\mu\nu} \right) + \mathcal{O}(\epsilon^2) \ , \nonumber
\end{eqnarray}
where in the first line we recognize the Einstein-Hilbert Lagrangian of GR with a cosmological constant term, while the second line encodes linear corrections in the EiBI parameter $\epsilon$ (and quadratic in curvature scalars).

Performing independent variations of the action (\ref{eq:BIaction}) with respect to metric and connection yields two sets of field equations

\begin{eqnarray}
\frac{\sqrt{-g}}{\sqrt{-q}} g^{\mu\nu}-\lambda g^{\mu\nu}&=&-\kappa^2 \epsilon T^{\mu\nu} \label{eq:metricfeq} \\
\nabla_{\alpha}\left(\sqrt{-q} q^{\mu\nu} \right)&=&0 \ , \label{eq:connectionfeq}
\end{eqnarray}
where $T_{\mu\nu}=\frac{2}{\sqrt{-g}} \frac{\delta \mathcal{S}_M}{\delta g^{\mu\nu}}$ (with $S_M=S_M(g_{\mu\nu},\psi_M)$ the action for the matter fields $\psi_M$) is the energy-momentum tensor of the matter, and we have defined the rank-two tensor $q_{\mu\nu}\equiv g_{\mu\nu}+\epsilon R_{\mu\nu}$\footnote{It should be noted that the physical content of such a new metric is related to the tensor perturbations (i.e. gravitational waves) on these backgrounds, see e.g. \cite{BHORgw}.}. Eq.(\ref{eq:connectionfeq}) implies that the independent connection $\Gamma_{\mu\nu}^{\lambda}$ can be solved as the Christoffel symbols of the metric $q_{\mu\nu}$, i.e.:
\begin{equation}
\Gamma_{\mu\nu}^{\lambda}=\frac{q^{\lambda\beta}}{2} \left(\partial_{\mu}q_{\nu\beta}+\partial_{\mu}q_{\nu\beta}-\partial_{\beta}q_{\mu\nu} \right) \ .
\end{equation}
The relation between the spacetime metric $g_{\mu\nu}$ and the \emph{auxiliary} metric $q_{\mu\nu}$ follows from the metric field equations (\ref{eq:metricfeq}) as
\begin{equation} \label{eq:qgom}
q_{\mu\nu}=g_{\mu\alpha}{\Omega^\alpha}_{\nu} \ ,
\end{equation}
where the object $\hat{\Omega}$ (in what follows a hat denotes a matrix) is defined as
\begin{equation} \label{eq:Omegadef}
\vert \hat{\Omega} \vert^{1/2} {(\hat{\Omega}^{-1})^\mu}_{\nu}=\lambda {\delta^\mu}_{\nu} - \epsilon \kappa^2 {T^\mu}_{\nu} \ ,
\end{equation}
from where it is clear that the transformation (\ref{eq:qgom}) between $g_{\mu\nu}$ and $q_{\mu\nu}$ depends only on the matter sources. Now, contracting (\ref{eq:metricfeq}) with $q^{\mu\alpha}$ and using the transformation (\ref{eq:qgom}) one finds the result
\begin{equation} \label{eq:Rmunuq}
{R^\mu}_{\nu}(q)=\frac{\kappa^2}{\vert \hat \Omega \vert^{1/2}} \left(\mathcal{L}_G {\delta^\mu}_{\nu} + {T^\mu}_{\nu} \right) \ ,
\end{equation}
where the gravitational Lagrangian, $\mathcal{L}_G$, turns out to be
\begin{equation}
\mathcal{L}_G=\frac{ \vert \hat{\Omega}  \vert^{1/2} - \lambda}{\epsilon \kappa^2} \ ,
\end{equation}
and ${R^\mu}_{\nu}(q)\equiv R^{\mu\alpha}q_{\alpha \nu}$. Eqs.(\ref{eq:Rmunuq}) represent a set of second-order, Einstein-like field equations for the $q_{\mu\nu}$ geometry, where all the contributions on the right-hand side are just functions of the matter sources and, as such, can be collectively read off as an effective energy-momentum tensor. This also means that, in vacuum, ${T_\mu}^{\nu}=0$, one has that $g_{\mu\nu}=q_{\mu\nu}$ (modulo a trivial re-scaling) and the solutions of the field equations (\ref{eq:Rmunuq}) correspond to those of General Relativity with an effective cosmological constant term $\Lambda_{eff}$, consistently with the statement above the expansion (\ref{eq:gsbi}). This guarantees the absence of ghost-like propagating degrees of freedom in this framework and, due to the fact that the spacetime metric $g_{\mu\nu}$ is related to the auxiliary metric $q_{\mu\nu}$ via the matter-mediated transformations (\ref{eq:qgom}), the field equations for $g_{\mu\nu}$ will be second-order and ghost-free as well. This is a rather generic property of metric-affine theories \cite{BHOR,Afonso:2017bxr}.

\subsection{Matter sector} \label{sec:matter}

The general form of the energy-momentum tensor of an anisotropic fluid (where we implicitly assume a spherically symmetric spacetime) is given by \cite{HS1997}

\begin{equation} \label{eq:Tmunufluid0}
{T^\mu}_\nu=(\rho + p_{\perp}) u^{\mu}u_{\nu}+p_{\perp} {\delta^\mu}_{\nu} + (p_r-p_{\perp}) \chi^{\mu}\chi_{\nu} \ ,
\end{equation}
where $u^\mu$ and $\chi^{\mu}$ represent normalized timelike and spacelike vectors, respectively,  such that $u^{\mu}\chi_{\mu}=0$. On the other hand $\rho(r)$ is the energy density of the fluid, $p_r(r)$ the pressure in the direction of $\chi_{\mu}$, and $p_{\perp}(r)$ the tangential pressure in the orthogonal direction to $\chi_{\mu}$. Note that in comoving coordinates the energy-momentum tensor (\ref{eq:Tmunufluid0}) can be cast under the more familiar form
\begin{equation} \label{eq:Tmunufluid2}
{T^\mu}_\nu=diag(-\rho,p_{r},p_{\perp},p_{\perp}) \ .
\end{equation}
In general, it is not possible to solve the field equations (\ref{eq:Rmunuq}) for an arbitrary shape of the density and pressure profiles of the fluid (not even in GR), so simplifying assumptions have to be made. As stated in the introduction, in this work we shall constraint these functions by demanding
\begin{enumerate}[i)]
\item the recovery of the Reissner-Nordstr\"om solution of the Einstein-Maxwell field equations far from the center,
\item the fulfillment of classical energy conditions,
\item correspondence with models of nonlinear electrodynamics.
\end{enumerate}

Regarding the last constraint, nonlinear electrodynamics have been frequently employed in gravitational scenarios in order to solve the singularity problem, though such attempts have been only partially successful, see e.g. \cite{AB} and the criticism of \cite{Bronnikov}. The energy-momentum tensor of the fluid (\ref{eq:Tmunufluid2}) can actually be mapped to that of nonlinear electrodynamics\footnote{Such models are defined in terms of a Lagrangian density of the form $\varphi(X,Y)$, where $X=\frac{1}{2}F_{\mu\nu}F^{\mu\nu}$ and $Y=\frac{1}{2}F_{\mu\nu}F^{*\mu\nu}$ are the two field invariants that can be built out the field strength tensor $F_{\mu\nu}=\partial_{\mu}A_{\nu}-\partial_{\nu}A_{\mu}$ and its dual $F^{*\mu\nu}=\frac{1}{2}\varepsilon^{\mu\nu\alpha\beta}F_{\alpha\beta}$. For electrostatic solutions one finds that $Y=0$.} if one chooses

\begin{equation} \label{eq:Tmunufluid}
{T^\mu}_\nu=diag(-\rho,-\rho,K(\rho),K(\rho)) \ ,
\end{equation}
where the function $K(\rho)$ thus characterizes both the fluid and nonlinear electrodynamics. To satisfy the other two constraints above on the fluid, a natural ansatz for the function $K(\rho)$ is that of

\begin{equation} \label{eq:Krho}
K(\rho)=\alpha \rho + \beta \rho^2 \ .
\end{equation}
When $\beta=0$, imposing equivalence of the energy-momentum tensor of the fluid and that of nonlinear electrodynamics yields the Lagrangian density $\varphi(X)=X^{\frac{1+\alpha}{2\alpha}}$. It should be stressed that this indeed was the case pursued in \cite{Shaikh2015}, where Lorentzian wormholes were found and characterized and which, in turn, is a generalization of the $\alpha=1$ case (corresponding to a standard Maxwell field $\varphi(X)=X$) studied in \cite{orRN}. As we want to recover the Reissner-Nordstr\"om solution of GR at large distances for our solutions, from now on we set $\lambda=1$ for asymptotic flatness and $\alpha=1$ (but $\beta \neq 0$) for a Maxwell behaviour at asymptotic infinity. This way, we shall let the new corrections encoded in the $\beta \rho^2$ terms in Eq.(\ref{eq:Krho}) to modify the geometry and we will study its properties.

To proceed further we first note that a standard conservation law for the matter fields holds in our scenario, $\nabla_{\mu}^{(g)}T^{\mu\nu}=0$\footnote{Note that $\nabla_{\mu}^{(g)}$ is the standard covariant derivative constructed with the Christoffel symbols of the spacetime metric $g_{\mu\nu}$. In general, in Palatini theories of gravity one has $\nabla_{\mu}^{(q)}T^{\mu\nu}\neq 0$, with $q_{\mu\nu}$ the auxiliary metric constructed with the independent connection. As far as the connection does not enter into the matter piece of the action (as is the present case), conservation of energy and momentum in these theories is automatically guaranteed.}. For a static, spherically symmetric line element of the form $ds^2=-C(x)dt^2+B^{-1}(x)dx^2+r^2(x)d\Omega^2$ (where $d\Omega^2=d\theta^2 + \sin^2(\theta) d\phi^2$ is the line element on the unit two-spheres) and for the ansatz (\ref{eq:Tmunufluid}), this conservation law reads explicitly
\begin{equation} \label{eq:coneq}
\rho_x+2[\rho+K(\rho)]\frac{r_x}{r}=0 \ ,
\end{equation}
where $\rho_x \equiv d\rho/dx$ and $r_x \equiv dr/dx$. Specifying the function $K(\rho)$ of Eq.(\ref{eq:Krho}) allows to integrate (\ref{eq:coneq}) as
\begin{equation} \label{eq:rhor}
\rho(r)=\frac{2\rho_0}{\left( \frac{r}{r_0} \right)^4-\beta \rho_0} \ ,
\end{equation}
where $r_0$ and $\rho_0$ are integration constants. To absorb these constants and simplify calculations it is useful to introduce a new (dimensionless) radial function $z=r/r_{\star}$, where $r_{\star}=r_0(\vert \beta \vert \rho_0)^{1/4}$. This way, the energy density of the fluid can be written under the compact form
\begin{equation} \label{eq:rhoz}
\rho(z)=\frac{\rho_m}{z^4-s_{\beta}} \ ,
\end{equation}
where $s_{\beta} \equiv \beta/\vert \beta \vert$ is the sign of $\beta$ and we have defined $\rho_m=2/\vert \beta \vert$. In these units, the asymptotic Maxwell limit is naturally achieved by identifying the electric charge as $Q^2=\kappa^2\rho_m r_{\star}^4$.

From the expression above it is clear that there are two different classes of behaviours for the energy density:
\begin{itemize}
\item For $s_{\beta}=+1$ it blows up at the finite radius $z=1$.
\item For $s_{\beta}=-1$ it reaches its maximum value $\rho=\rho_m$ at the radius $z=0$.
\end{itemize}
It should be noted that both these two branches of solutions satisfy the weak energy condition. Indeed, the first half of such a condition states that $\rho + p_{r} \geq 0$, which for (\ref{eq:Tmunufluid}) is trivially fulfilled, while the second half, $\rho + p_{\theta} \geq 0$ and $\rho + p_{\varphi} \geq 0$, for the choice (\ref{eq:Krho}) implies that $\rho_m +s_{\beta}  \rho\geq 0$. Thus, for $s_{\beta}=+1$ this is trivially satisfied, while for $s_{\beta}=-1$ it is also satisfied due to the presence of the bound $\rho \leq \rho_m$.

With these constraints now the field equations (\ref{eq:Rmunuq}) can be cast in amenable form for calculations. First, given that the deformation matrix (\ref{eq:Omegadef}) is determined by the energy-momentum tensor, the algebraic structure of the latter defined in (\ref{eq:Tmunufluid}) in two $2\times 2$ blocks allows to consistently introduce the ansatz for the matrix $\hat{\Omega}$ as
\begin{equation} \label{eq:Omegafluid}
\hat{\Omega}=
\left(\begin{array}{lr} \Omega_1 I_{2\times2} & 0_{2\times2} \\ 0_{2\times2} & \Omega_2 I_{2\times2} \end{array}\right) \ ,
\end{equation}
where $I_{2\times2}$ and $0_{2\times2}$ are the $2\times2$ identity and zero matrices, respectively, while consistency with Eq.(\ref{eq:Omegadef}) tells us that
\begin{equation}
\Omega_{1}=1-\kappa^2 \epsilon K(\rho)
\hspace{0.1cm};\hspace{0.1cm}
\Omega_{2}=1 + \kappa^2 \epsilon \rho \ .
\end{equation}
Now it is a matter of just a little algebra to show that the field equations (\ref{eq:Rmunuq}) become
\begin{equation} \label{eq:Rmunufluid}
{R^\mu}_{\nu}(q)=\frac{1}{\epsilon}
\left(\begin{array}{lr} \left(\frac{\Omega_1-1}{\Omega_1}\right) I_{2\times2} & 0_{2\times2} \\ 0_{2\times2} & \left(\frac{\Omega_2-1}{\Omega_2}\right) I_{2\times2} \end{array}\right) \ .
\end{equation}
and they are now ready for their resolution.

\section{Solution of the field equations} \label{sec:III}

To solve the field equations (\ref{eq:Rmunufluid}) we first introduce a static, spherically symmetric line element for the auxiliary geometry $q_{\mu\nu}$ as
\begin{equation} \label{eq:lineqgen}
ds_q^2=-e^{2\psi(x)}A(x)dt^2+\frac{1}{A(x)}dx^2+ x^2 d\Omega^2 \ .
\end{equation}
Using the symmetry of the fluid energy-momentum tensor (\ref{eq:Tmunufluid}), ${T^t}_t={T^x}_x$, from the computation of the components of the Ricci tensor it follows that the combination ${R^t}_t={R^x}_x$ in the field equations (\ref{eq:Rmunufluid}) yields $\psi=$constant, which can be set to zero by a redefinition of the time coordinate, without loss of generality. Now, introducing a standard mass ansatz as
\begin{equation} \label{eq:massf}
A(x)=1-\frac{2M(x)}{x} \ ,
\end{equation}
the component ${R^\theta}_\theta$ of the field equations (\ref{eq:Omegafluid}) yields the equation
\begin{equation} \label{eq:massx}
M_x=\frac{x^2}{2\epsilon} \frac{\Omega_2-1}{\Omega_2} \ .
\end{equation}
For the next step, let us introduce a line element for the spacetime metric $g_{\mu\nu}$ as
\begin{equation} \label{eq:lineggen}
\frac{ds^2}{r_{\star}^2}=g_{tt}dt^2+g_{xx}dx^2+z^2(x)d\Omega^2 \ ,
\end{equation}
where the notation $z=r/r_{\star}$ is the same as that introduced for the fluid in section \ref{sec:matter}. In order not to overload the notation, from now on we will bear in mind that the coordinates $t$ and $x$ are also expressed in units of $r_{\star}$. From the relation (\ref{eq:qgom}) with the structure (\ref{eq:Omegafluid}), we obtain the relation between the radial functions in the spacetime (\ref{eq:lineggen}) and auxiliary (\ref{eq:lineqgen}) geometries as
\begin{equation} \label{eq:xz}
x^2=z^2 \Omega_2 \ .
\end{equation}
This relation will be very important later when characterizing the different solutions. But before going into that, let us keep solving the field equations, for which we take a derivative upon (\ref{eq:xz}) and using the continuity equation of the fluid (\ref{eq:coneq}) one arrives to the result
\begin{equation} \label{eq:dxdza}
\frac{dx}{dz}=\Omega_2^{1/2} \left[1-\frac{\epsilon \kappa^2}{\Omega_2} \left(\rho + K(\rho) \right) \right] \ .
\end{equation}
This allows to write Eq.(\ref{eq:massx}) as
\begin{equation}
\frac{dM}{dz}=r_{\star}^3\frac{z^2 \Omega_2^{1/2} (\Omega_2-1)}{2\epsilon} \left[1-\frac{\epsilon \kappa^2}{\Omega_2} \left(\rho + K(\rho) \right) \right] \ .
\end{equation}
Now, we formally write the integration of this function as $M(z)=M_0(1+\delta_1 G(z))$, where $M_0$ is the Schwarzschild mass, $G(z)$ contains the fluid contribution, and all the constants have been isolated in $\delta_1$. After playing all these tricks, and taking the form of $K(\rho)$ specified in (\ref{eq:Krho}), the line element for the spacetime metric (\ref{eq:lineggen}) can be conveniently written as
\begin{equation} \label{eq:linefinal}
\frac{ds^2}{r_\star^2}=-\frac{A(x)}{\Omega_1}dt^2 + \frac{dx^2}{A(x)\Omega_1} + z^2(x)d\Omega^2 \ ,
\end{equation}
with the compact expressions
\begin{eqnarray}
A(z)&=&1-\frac{r_S (1+\delta_1 G(z)) }{r_{\star}z\Omega_2^{1/2}}  \label{eq:Afunction}\\
\delta_1&=& \frac{  r_{\star}^3}{r_S l_m^2}\label{eq:delta1}\\
\Omega_1&=&1-s_{\epsilon} \xi^2 \left( \frac{z^4+s_{\beta}}{(z^4-s_{\beta})^2} \right)\\
\Omega_2&=&1+\frac{s_{\epsilon} \xi^2}{z^4-s_{\beta}} \\
G_z&\equiv&\frac{dG}{dz}= \frac{z^2 \Omega_1}{ (z^4-s_{\beta})\Omega_2^{1/2}} \ , \label{eq:Gz}
\end{eqnarray}
where $r_S=2M_0$ is the Schwarzschild radius and we have introduced the new scale $\xi^2 \equiv l_{\epsilon}^2/l_{m}^2$, with $l_{\epsilon}^2= \vert \epsilon \vert$ and $l_{m}^2=(\kappa^2 \rho_m)^{-1}$. Note that the transformation (\ref{eq:dxdza}) between the two systems of coordinates can be written as
\begin{equation} \label{eq:dxdrg}
\frac{dx}{dz}=\frac{\Omega_1}{\Omega_{2}^{1/2}} \ ,
\end{equation}
which will be very useful later. The line element (\ref{eq:linefinal}) together with the definitions above is the master set of equations that we will use in Sec. \ref{sec:V} to study the properties of the corresponding solutions. But before going into that, let us have a look at the geodesic equations in these theories.

\section{Geodesic structure} \label{sec:geo}

For the sake of the discussion below on the properties of the different classes of configurations, let us introduce here the main elements for the analysis of the geodesic behaviour in the corresponding theories. Given a geodesic curve $\gamma^{\mu}=x^{\mu}(u)$, where $u$ is the affine parameter, in a coordinate basis the geodesic equation can be written as \cite{Chandra}
\begin{equation} \label{eq:geoeq}
\frac{d^2x^{\mu}}{du^2} + \Gamma^{\mu}_{\alpha\beta}\frac{dx^{\alpha}}{du} \frac{dx^{\beta}}{du} =0 \ ,
\end{equation}
which is a second-order differential equation to be supplied with initial conditions $x^{\mu}(0)$ and $dx^{\mu}/du\vert_0$. The general formalism for geodesic motion in Palatini theories of gravity has been developed with certain detail in \cite{OlmoBook}. First thing to note is that the matter sector of our theory, as described by the energy-momentum tensor (\ref{eq:Tmunufluid}), is assumed to couple to the gravitational sector (\ref{eq:gsbi}) only via the metric and the matter fields (and not via the connection). This implies that photons and free-falling particles will follow geodesics of the spacetime metric $g_{\mu\nu}$ in Eq.(\ref{eq:geoeq}), in compliance with Einstein's equivalence principle\footnote{Should one allow a coupling of the matter sector with the independent connection $\Gamma_{\mu\nu}^{\lambda}$, then one would need to regard geodesics of the auxiliary metric $q_{\mu\nu}$ as physically meaningful.}. Second, due to the spherical symmetry of our problem, we can rotate the plane of motion to make it coincide with $\theta=\pi/2$, without loss of generality, and, furthermore, we can introduce two conserved quantities of motion, $E=Bdt/d\lambda$ and $L=r^2 d\theta/d\lambda$, where $B=A/\Omega_{1}$. For time-like observers, $u^{\mu}u_{\mu}=-1$, these quantities can be interpreted as the particle's energy and angular momentum per unit mass, respectively. For null geodesics, $u^{\mu}u_{\mu}=0$, this interpretation cannot be sustained, but the quotient $L/E$ can be identified instead as an apparent impact parameter as seen from asymptotic infinity.

After all these considerations, the geodesic equation for a geometry of the form (\ref{eq:linefinal}) can be written as \cite{OlmoBook}
\begin{equation} \label{eq:geoBI}
\frac{1}{\Omega_{1}^2} \left(\frac{dx}{du}\right)^2=E^2-V_{eff} \ ,
\end{equation}
where the effective potential $V_{eff}$ takes the form
\begin{equation} \label{eq:Veff}
V_{eff}=B\left(\frac{L^2}{r^2(x)} - \kappa \right) \ ,
\end{equation}
with $\kappa=0$ for null geodesics and $\kappa=-1$ for time-like particles. Introducing the simple change of coordinates $dy=dx/\Omega_1$, then Eq.(\ref{eq:geoBI}) becomes a single differential equation akin to the movement of a one-dimensional particle in the effective potential $V_{eff}$, which facilitates its resolution, as we shall see in the different cases studied in next section.

\section{Analysis of the solutions} \label{sec:V}

\subsection{Radial function} \label{eq:RadFun}

The relation (\ref{eq:xz}) between the radial functions in the auxiliary and spacetime geometries can be explicitly written as
\begin{equation}  \label{eq:xzexplicit}
x^2=z^2\left(1+\frac{s_{\epsilon} \xi^2}{z^4-s_{\beta}} \right) \ .
\end{equation}
This can be expressed as a cubic equation for the variable $z^2$ as
\begin{equation} \label{eq:whcubic}
(z^2)^3-x^2(z^2)^2+(s_{\epsilon} \xi^2-s_{\beta})(z^2)+s_{\beta}x^2=0 \ .
\end{equation}
Though this equation admits a (cumbersome) analytical solution,  we find it more convenient to discuss the relevant cases by direct inspection of the relation (\ref{eq:xzexplicit}). This yields a natural classification in terms of four different  configurations:
\begin{itemize}
\item Case I: For $\{s_{\epsilon}=-1,s_{\beta}=+1\}$, the radial function $z$ reaches a minimum at $z_{c}=(1 + \xi^2)^{1/4}$, where $x=0$ and the density is finite [see Eq.(\ref{eq:rhoz})]. At this point the radial function $z(x)$ bounces off and re-expands again. This bouncing behaviour signals the existence of a \emph{wormhole}, a topologically non-trivial structure connecting two asymptotically flat regions of the spacetime \cite{Visser}, with $z_{c}$ representing its throat (further details will be provided in section \ref{sec:CaseI} below). Thus, in this case, one needs two copies of the radial function $z \in ( z_{c},\infty)$ to cover the whole manifold, or a single chart when using $x \in (-\infty,+\infty)$.
\item Case II: For $\{s_{\epsilon}=-1,s_{\beta}=-1\}$, there are two classes of configurations separated by the threshold $\xi^2=1$. In this sense, for $\xi^2>1$, the value $x=0$ is attained at a minimum radius $z_{c}^4= \xi^2 -1$, while for $\xi^2<1$ one finds $x=0$ at $z=0$. The wormhole interpretation is natural for the former (for which $\rho<\rho_m$), but dubious for the latter.
\item Case III: For $\{s_{\epsilon}=+1,s_{\beta}=-1\}$, as $x \rightarrow 0$ one finds that $z \rightarrow 0$ too. A bouncing behaviour for $z(x)$ arises again, though now the transition between the two regions $x \in (0,+\infty)$ and $x  \in (-\infty,0)$ is not smooth.
\item Case IV: For $\{s_{\epsilon}=+1,s_{\beta}=+1\}$, as the radial function $z \rightarrow 1$ (its minimum value) one finds $x \rightarrow \infty$.
\end{itemize}

In what follows we shall split our analysis into the four cases above to study separately their properties.

\subsection{Case I: $\{s_{\epsilon}=-1,s_{\beta}=+1\}$} \label{sec:CaseI}

For this case, the relevant functions characterizing the matter and the geometry (\ref{eq:linefinal}) take the form
\begin{eqnarray}
\rho&=&\frac{\rho_m}{z^4-1} \label{eq:CIed} \\
\Omega_{1} &=& 1 + \frac{\xi^2(z^4+1)}{(z^4-1)^2} \hspace{0.1cm};\hspace{0.1cm} \Omega_{2}= 1-\frac{\xi^2}{z^4-1} \\
G_z&=& \frac{z^2 \Omega_1}{(z^4-1)\Omega_2^{1/2}}
\end{eqnarray}
The function $G_z$ admits an exact analytical integration given by
\begin{eqnarray}
G(z)&=&  \frac{2  \Bigg( \frac{z_c^4}{z^4} F_1\Big(\frac{5}{4};\frac{1}{2},\frac{1}{2};\frac{9}{4};\frac{1}{z^4},\frac{z_c^4}{z^4}\Big)}{15 \xi ^2  z } \nonumber   \\
&-& \frac{ 5 (2 \xi ^2+1) F_1\Big(\frac{1}{4};\frac{1}{2},\frac{1}{2};\frac{5}{4};\frac{1}{z^4},\frac{z_c^4}{z^4}\Big) \Bigg) } {15 \xi ^2  z}  \label{eq:GcaseI}   \\
&+& \frac{ 5z^3 (z^4-z_c^4)^{3/2} (-\xi ^2+\frac{2}{z^4}-2)}{15 \xi ^2 (z^4-1)^{3/2}} \nonumber
\end{eqnarray}
where $z_c=(\xi^2+1)^{1/4}$ is the minimum radius of the radial function, and $F_1[a,b_1,b_2,c,x,y]$ is the Appell hypergeometric function of two variables $(x,y)$. The above function reproduces the expected GR behaviour at large distances $z \rightarrow \infty$, namely, $G(z) \simeq 1/z^2 + \mathcal{O}(1/z^6)$ and $G(z) \simeq -1/z + \mathcal{O}(1/z^5)$. Note that Eq.(\ref{eq:GcaseI}) would allow to obtain closed expressions for the metric functions, though in cumbersome and not too illuminating forms, so we shall not explicitly write them here. Nonetheless we can check that for large distances, $z \gg 1$, these functions become
\begin{eqnarray}
g_{tt} &\approx& - \left(1-\frac{r_S}{r_{\star} z} +\frac{r_S \delta_1}{r_{\star} z^2} \right) + \frac{\xi^2}{z^4} + \mathcal{O}\left(\frac{\xi^2}{z^5} \right) \label{eq:gttinfI}  \\
g_{rr} &\approx& \left(1-\frac{r_S}{r_{\star} z} +\frac{r_S \delta_1}{r_{\star} z^2} + \frac{\xi^2}{z^4} + \mathcal{O}\left(\frac{\xi^2}{z^5} \right)  \right)^{-1}
\end{eqnarray}
which, after restoring the notation, is nothing but the Reissner-Nordstr\"om solution of GR, $g_{tt}=g_{rr}^{-1}=1-r_S/r+Q^2/r^2$ (plus $\xi^2$-corrections), in agreement with the recovery of Maxwell Lagrangian in the asymptotic limit of the matter sector. This is a shared feature for all the solutions obtained in this work (Cases II, III and IV below).

\begin{figure}[h]
\centering
\includegraphics[width=0.50\textwidth]{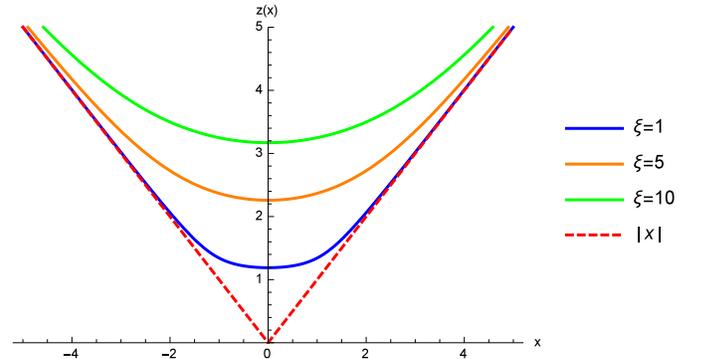}
\caption{Radial function $z(x)$ for the Case I, as follows from integration of Eq.(\ref{eq:whcubic}) in this case. From bottom to top the solid curves represent $\xi=1$ (blue), $\xi=5$ (orange) and $\xi=10$ (green), with the dashed red line corresponding to $\vert x \vert $. The wormhole throat is located at $z_c=(\xi^2+1)^{1/4}$. \label{fig:whcaseI}}
\end{figure}

We are mostly interested in the modifications on the structure of these solutions as compared to the Reissner-Nordstr\"om one, which become significant only in the innermost region. We already know that the minimum value attained by the radial function corresponds to $z_c=(\xi^2+1)^{1/4}$ where it bounces off, which allows to infer the presence of a wormhole structure with $z_c$ representing its throat. In Fig.\ref{fig:whcaseI} we have depicted this structure, where we show the growth of the size of the throat as $\xi^2$ is increased.

As $z_c>1$, this means that the energy density of the fluid in this case, as given by Eq.(\ref{eq:CIed}), will always be bounded. Now, expanding the relevant metric functions around $z=z_c$, one finds
\begin{eqnarray}
\Omega_{1} &\approx& \frac{2z_c^4}{z_c^4-1} -\frac{4z_c^3(z_c^4+3)}{(z_c^4-1)^2}(z-z_c)+\mathcal{O}(z-z_c)^2 \label{eq:Omega1a} \\
\Omega_{2} &\approx& \frac{4z_c^3}{z_c^4-1} (z-z_c) + \mathcal{O}(z-z_c)^2 \label{eq:Omega2a} \\
z&\approx& z_c +\left(\frac{z_c^4-1}{4z_c^5}\right) x^2 \label{eq:radialcaseI} \\
G(z)&\approx& -\frac{1}{\delta_c} +2C_1 (z-z_c)^{1/2} + \mathcal{O}(z-z_c)^{3/2} \ , \label{eq:GzcaseA}
\end{eqnarray}
where for convenience we have introduced the constant $C_1=\left(\frac{z_c^3}{z_c^4-1} \right)^{3/2}$, while we have another constant
\begin{equation} \label{eq:deltacI}
\delta_c=-\frac{\xi ^2 \Gamma (-\frac{1}{4}) \Gamma (\frac{7}{4})}{\sqrt{2} \pi ^{3/2} z_c^3 \, _2F_1(-\frac{3}{4},\frac{1}{2};\frac{3}{4};\frac{1}{z_c^4})} >0 \ ,
\end{equation}
(where $\Gamma[a]$ is Euler's gamma function) whose explicit value comes from requiring the matching of the asymptotic and inner expansions of the metric functions. This constant plays a key role in the characterization of the solutions, as shall be shown below. Note that the expression of the radial function around the wormhole throat in Eq.(\ref{eq:radialcaseI}) is consistent with the bouncing behaviour depicted in Fig.\ref{fig:whcaseI}.

\begin{figure}[h]
\centering
\includegraphics[width=0.45\textwidth]{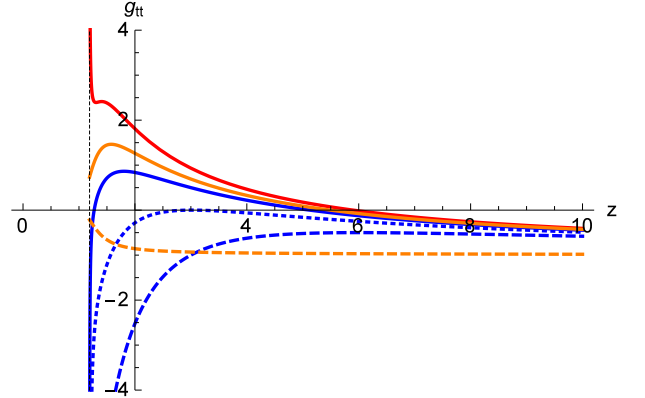}
\caption{Metric component $g_{tt}(z)$ of Case I taking $\xi=1$, for which the wormhole throat is located at $z_{c}=2^{1/4}$ (represented by the vertical dashed black line). We find i) Reissner-Nordstr\"om-like solutions with two (blue solid, $\delta_1=1/10,r_{\star}/r_S=1/6$), a single degenerate (blue dotted, $\delta_1=3,r_{\star}/r_S=1/6$) or zero (blue dashed, $\delta_1=3/4, r_{\star}/r_S=1/6$) horizons; ii) Schwarzschild-like solutions with a single horizon (red solid, $\delta_1=9/6,r_{\star}/r_S=1/6$) and iii) Minkowski-like solutions with a single horizon (orange solid, $\delta_1=\delta_c \approx 0.464 ,\delta_2=5$) or none (orange dashed, $\delta_1=\delta_c \approx 0.464 ,r_{\star}/r_S=1/6$). All solutions are asymptotically flat. \label{fig:2}}
\end{figure}

Now, expanding the metric components $g_{tt}$ and $g_{rr}$ around $z=z_c$ yields the result
\begin{eqnarray}
g_{tt} &\approx& -\frac{r_S (\delta_1/\delta_c-1)}{4r_{\star} z_c^2 C_1\sqrt{z-z_c}} -\frac{1}{2z_c C_1^{2/3}} \left(1-\frac{r_S C_1^{2/3} \delta_1}{r_{\star} z_c} \right) \nonumber \\
&+& \mathcal{O}\left(\sqrt{z-z_c}\right) \label{eq:gttCaseI} \\
g_{rr} &\approx& \frac{r_{\star}z_c^2}{r_S  C_1^{1/3} (\delta_1/\delta_c -1) \sqrt{z-z_c}} + \mathcal{O}\left(1\right) \label{eq:gttcaseII}
\end{eqnarray}
which shows that, in general, the metric component $g_{tt}$ is divergent there, the sign  being controlled by the ratio $\delta_1/\delta_c$. On the contrary, for $\delta_1=\delta_c$, the first term in the expansion vanishes and, therefore, $g_{tt}$ becomes finite at the wormhole throat. These expressions have a non-trivial impact on the causal structure of the corresponding geometries. Indeed, as depicted in Fig. \ref{fig:2}, several classes of configurations may be found. In this sense, for $\delta_1/\delta_c>1$ one finds the presence of Reissner-Nordstr\"om-type solutions, with two horizons, a single but degenerate one (corresponding to extreme black holes) or no horizons, while for $\delta_1/\delta_c<1$ a Schwarzschild-like black hole arises instead, characterized by a single non-degenerate horizon (recall that the horizons are located symmetrically on each side of the horizon).  On the other hand, for $\delta_1 = \delta_c$ one finds either black holes with a single horizon or none, depending on the particular values of the parameters characterizing the solutions. This structure of horizons is generic for any value of the typical scale of the theory (encoded in $\xi^2$). Moreover it exactly matches the typical structure of Born-Infeld black holes in GR \cite{BI-GR} and, more generally, of those GR black holes supported by nonlinear electromagnetic fields whose electrostatic configurations attain a maximum value at the center \cite{NEDg-GR}.

To further understand the innermost structure of these solutions let us consider the behaviour of the Kretchsman scalar, $K={R_\alpha}^{\beta\mu\nu}{R^\alpha}_{\beta\mu\nu}$. For large distances, $z \gg z_c$, one gets
\begin{equation}
K \approx \frac{12}{\delta_2^2 z^6} -\frac{48 \delta_1}{\delta_2^2 z^7} + \frac{56 \delta_1^2}{\delta_2^2 z^8} + \frac{72 \xi^2}{\delta_2 z^9} + \mathcal{O}\left(\frac{\xi^2}{z^{10}}\right) \ ,
\end{equation}
where the first three terms correspond to the expected behaviour of the Reissner-Nordstr\"om solution of GR, in agreement with the recovery of that solution in the asymptotic limit. On the other hand, at the wormhole throat, $z=z_c$, one finds an expression that can be arranged under the following form:
\begin{eqnarray}
K &\approx& \frac{(\delta_1-\delta_c)}{(z-z_c)^3}\left(\frac{r_s^2 (\delta_1-\delta_c)}{4r_{\star}^2\delta_c^2z_c^4C_1^{2/3}} +\mathcal{O}(z-z_c)\right) \\
&+& a + \mathcal{O}(z-z_c) \nonumber
\end{eqnarray}
where $a=a(r_S,z_c,r_{\star},\delta_c,\delta_1)$ is a constant with an involved dependence on the model and solution parameters. Let us note that the leading-order divergence in this expression has been softened down to $\sim 1/(z-z_c)^{3}$ as compared to the GR result. Moreover, when $\delta_1=\delta_c$, replacing first this choice in the metric function and expanding next the Kretchsman scalar around the wormhole throat yields the finite result $K=a(r_S,z_c,r_{\star},\delta_c)+\mathcal{O}(z-z_c)$ (and similarly for other curvature invariants such as the Ricci scalar $g_{\mu\nu}R^{\mu\nu}$ or the Ricci-squared $R_{\mu\nu}R^{\mu\nu}$), which means that the Minkowskian solutions are free of curvature divergences everywhere\footnote{Let us recall that, from the discussion of the horizons above, these finite-curvature solutions may be cloaked by an event horizon or be naked instead.}. This is in contrast with the results found in the case of GR coupled to nonlinear electrodynamics. In those cases, despite the existence of electrostatic solutions with finite curvature scalars \cite{AB}, this cannot be achieved via models defined as a single-branch function satisfying standard energy conditions, see \cite{Bronnikov}.

\begin{figure}[h]
\centering
\includegraphics[width=0.50\textwidth]{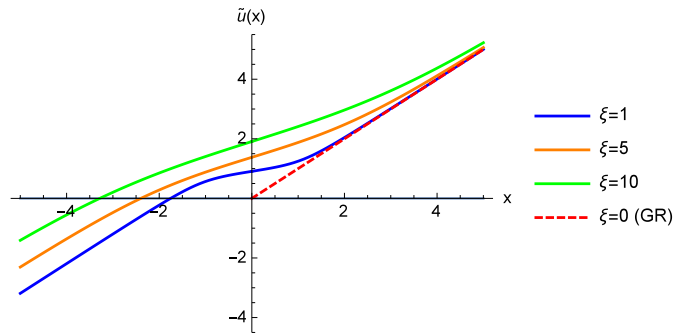}
\caption{The affine parameter $\tilde{u}(x)$ for null radial geodesics of Case I, as given by Eq.(\ref{eq:nullradial2}). Here we take values $\xi=1$ (blue), $\xi=5$ (orange) and $\xi=10$ (green), with the dashed red curve representing $\tilde{u}(x)=x$ and corresponding to the GR behaviour. As it is obvious from this plot, null radial geodesics in this case are complete. \label{fig:geonullI}}
\end{figure}

Let us now consider the implications of the above results for the (in)completeness of geodesics, using the elements introduced in Sec. \ref{sec:geo}. The geodesic equation (\ref{eq:geoBI}), for null ($\kappa=0$) radial ($L=0$) geodesics, can be conveniently rewritten, using Eq.(\ref{eq:dxdrg}), as
\begin{equation}
\pm  E \cdot d\tilde{u}(x)= \frac{dz}{\Omega_2^{1/2}} \ ,
\end{equation}
where we have re-scaled $\tilde{u}(x) \equiv u(x)/r_{\star}$ and the sign $\pm$ corresponds to outgoing/ingoing geodesics, respectively (as seen from the $x>0$ side of the wormhole). This equation admits an analytic integration of the form
\begin{equation}\label{eq:nullradial2}
\pm E \cdot \tilde{u}(x)=
\left\{ \begin{tabular}{lr} $\zeta(z;\xi)$ & \text{ if } $x\ge 0$ \\
\text{ }\\
$2x_0(\xi)-\zeta(z;\xi)$ & \text{ if } $x\le 0$
\end{tabular} \right. \ ,
\end{equation}
where we have introduced the function
\begin{eqnarray}
\zeta(z;\xi)&=&\frac{\sqrt{(z^4-z_c^4)(z^4-1)}}{z^3} \nonumber \\
&+&\frac{1}{21z^3} \Bigg[7(3+\xi^2)F_1\left(\frac{3}{4},\frac{1}{2},\frac{1}{2},\frac{7}{4},\frac{1}{z^4},\frac{z_c^4}{z^4}\right) \\
&-&\frac{9z_c^4}{z^4} F_1\left(\frac{7}{4},\frac{1}{2},\frac{1}{2},\frac{11}{4},\frac{1}{z^4},\frac{z_c^4}{z^4}\right) \Bigg] \nonumber \ ,
\end{eqnarray}
and the set of constants
\begin{equation}
x_0(\xi)=\frac{5\pi^{3/2}\xi^2\Big(2 {_2}F_1\Big(\frac{3}{4},\frac{3}{2},\frac{5}{4}, \frac{1}{z_c^4} \Big) -  {_2}F_1\Big(\frac{1}{2},\frac{3}{4},\frac{5}{4}, \frac{1}{z_c^4} \Big)  \Big)}{32\sqrt{2}z_c^3\Gamma\left[\frac{5}{4}\right]\Gamma\left[\frac{9}{4}\right]} \ .
\end{equation}
where ${_2}F_1[a,b,c,z]$ is a hypergeometric function. The behaviour of these geodesics is depicted in Fig. \ref{fig:geonullI} for several values of the scale $\xi^2$. For $z \gg z_c $ one finds $\pm E \tilde{u}(x) \approx z+\mathcal{O}(z^{-3}) \approx x$ and one recovers the standard GR behaviour there. However, as one approaches the wormhole throat, $x=0$, one finds instead $\pm E (\tilde{u}(x)- \tilde{u}(0))\approx  \frac{2(z_c^4-1)^{1/2}}{z_c^{3/2}} \sqrt{z-z_c} =\left(\frac{z_c^4-1}{z_c^4}\right)x$. This behaviour allows each geodesic to be smoothly  extended across the wormhole throat to reach arbitrarily large values of its affine parameter. This is in sharp contrast with the GR behaviour, where the geodesic equation in that case, $dr/du=\pm E^2$, has the solution (for outgoing/ingoing geodesics) $\pm E u(r)=r$. Thus, in the GR case, as the function $r(u)$ is positive definite, the affine parameter is only defined on the positive/negative axis and these geodesics are incomplete. This shows that the presence of a wormhole structure in our case makes it possible to obtain complete null radial geodesics no matter the value of the scale $\xi^2$. Moreover, this result holds true despite the generic presence of curvature divergences at the wormhole throat. This follows from the fact that  only the case $\delta_1 = \delta_c$ is free of divergences, but the structure of the geodesics is insensitive to the value of $\delta_1$.

For time-like and null geodesics with $L \neq 0$ we need to turn our attention to the behaviour of the effective potential (\ref{eq:Veff}) in the geodesic equation (\ref{eq:geoBI}). From the expansions above of the metric functions, it follows that for large distances, $x \rightarrow \infty$, from Eq.(\ref{eq:gttinfI}) the potential behaves as $V_{eff} \approx \left(L^2/x^2-\kappa \right)$, which is nothing but the standard (positive, and negligible for our purposes) potential barrier of the Reissner-Nordstr\"om solution of GR. As we approach the wormhole throat, $x=0$, using (\ref{eq:radialcaseI}) this barrier is replaced there by
\begin{equation} \label{eq:effpot}
V_{eff} \approx -\frac{a}{\vert x \vert} - b +\mathcal{O}(x) \ ,
\end{equation}
where we have introduced the constants
\begin{eqnarray}
a&=&\frac{\xi^4}{2z_c^6} \frac{(\delta_c-\delta_1)}{\delta_c \delta_2}\left(\frac{L^2}{r_{\star}^2 z_c^2}- \kappa  \right) \label{eq:acaseI} \\
b&=&\frac{\xi^2}{2z_c^4}\frac{(\delta_2-\delta_1)}{\delta_2} \left(\frac{L^2}{r_{\star}^2 z_c^2}  - \kappa  \right) \ ,  \label{eq:bcaseI}
\end{eqnarray}
and defined $\delta_2=\frac{\xi^2 r_{\star}}{r_S z_c^2}$. This way we have reduced the problem for these geodesics to inspect the nature of the effective potential around the wormhole throat. There are three cases to be considered separately:

\begin{itemize}
\item $\delta_1>\delta_c$ (Reissner-Nordst\"om-like solutions): In this case one finds an infinite potential barrier as the wormhole throat is approached and, consequently, all geodesics bounce at some $z>z_c$ and remain in the $x>0$ region. Thus, in much the same way as all timelike and null geodesics with $L \neq 0$ of the Reissner-Nordstr\"om solution of GR, these geodesics are not able to reach the wormhole throat, being complete.
\item $\delta_1<\delta_c$ (Schwarzschild-like solutions): Now the potential changes from infinitely repulsive to infinitely attractive and, consequently, all these geodesics are unavoidably dragged towards the wormhole throat (depending on the combination of constants the effective potential could have a maximum and, in such cases, only geodesics whose energy $E$ is larger than it will get to the wormhole throat). With the approximate form of the effective potential as $x \rightarrow 0$, Eq.(\ref{eq:effpot}), one finds that the geodesic equation (\ref{eq:geoBI}) behaves in this region as
    \begin{equation}
    \frac{d\tilde{u}}{dx}=\frac{\xi^2}{2a^{1/2}(1+\xi^2)} \vert x \vert^{1/2} - \frac{\xi^2(b+E^2)}{4a^{3/2}(1+\xi^2)} \vert x \vert^{3/2} + \mathcal{O}(x^{5/2}) \ ,
    \end{equation}
    whose integration yields the result
    \begin{equation}
    \tilde{u}(x)=\frac{\xi^2}{3(1+\xi^2)} x \Big\vert \frac{x}{a} \Big\vert^{1/2} \left(1-\frac{3(b+E^2)}{10} \Big\vert \frac{x}{a} \Big\vert \right) + \mathcal{O}(x^{7/2}) \ .
    \end{equation}
    As the coordinate $x$ extends over the whole real axis, it is clear that these geodesics are complete for all values of the parameter $\xi^2$ and the other constants characterizing the solutions. This is so despite the divergence of both the effective potential and the curvature scalars as the wormhole throat is approached. Likewise the null radial case, such geodesics in GR (for Schwarzschild black holes) are incomplete due to the fact that $r=0$ is reached in finite affine time, with no possibility of further extension, a result avoided in this case thanks to the presence of the wormhole structure.
\item $\delta_1=\delta_c$ (Minkowski-like solutions): In this case the effective potential has a shape at the wormhole throat of the form: $V_{eff} \approx -b + c(\xi) x^2$ (with $c$ some constant with an involved dependence on $\xi^2$), which is finite there. Moreover, depending on $\xi^2$ and on the model parameters, there may be both minima and maxima, thus allowing for the existence of bounded orbits below the maximum. On the other hand, those particles with energies above the maximum of the potential will be able to reach the wormhole throat, with their affine parameter behaving there as
    \begin{equation}
    \tilde{u}(x)= \frac{\xi^2}{2(1+\xi^2)\sqrt{b+E^2}} x \left(1 +\frac{(\xi^2+4)}{6(1+\xi^2)^{3/2}} x^2\right)+ \mathcal{O}(x^5) \ .
    \end{equation}
    Again, due to the definition of the coordinate $x$ over the whole real axis, these geodesics can be naturally extended beyond the $x=0$ region, which implies their completeness.
\end{itemize}

Thus, we conclude that these geometries are null and timelike geodesically complete for all the spectrum of parameters characterizing the solutions. Since, in particular, the parameter $\delta_1$ contains the information about the number and type of horizons, this implies the existence of naked geodesically complete configurations, whose implications regarding the issue of regular black hole remnants are still to be investigated. On the other hand, the existence of curvature divergences at the wormhole throat, absent only when $\delta_1=\delta_c$, does not prevent in any way the extension of geodesics across the wormhole throat, as the affine parameter can be indefinitely continued. Since geodesics represent idealized point-like observers, there is still the question about the meaning and implications of such curvature divergences acting upon extended observers crossing the $x=0$ region. This has been explored in the case of EiBI gravity coupled to an electromagnetic (Maxwell) field in \cite{ors16}, where an analysis upon similar wormhole structures as those found here supports the view that no destructive effects would take place on observers crossing the throat.

\subsection{Case II: $\{s_{\epsilon}=-1,s_{\beta}=-1\}$} \label{sec:CaseII}

In this case we have the expressions
\begin{eqnarray}
\rho&=&\frac{\rho_m}{z^4+1} \label{eq:rhocII} \\
\Omega_1&=&1+\frac{\xi^2(z^4-1)}{(z^4+1)^2} \hspace{0.1cm} ; \hspace{0.1cm} \Omega_{2}=1-\frac{\xi^2}{z^4+1} \label{eq:OmcaseIIapp} \\
G_z&=&\frac{z^4 \Omega_1}{(z^4+1)\Omega_2^{1/2}} \ .
\end{eqnarray}
Now the function $\Omega_{2}$ vanishes at $z_c=(\xi^2-1)^{1/4}$, which sets a critical value for $\xi^2=1$, while the energy density of the matter fields (\ref{eq:rhocII}) is always finite. The analysis now needs to be split into three subcases.

\subsubsection{$\xi^2>1$} \label{sec:CaseIIa}

\begin{figure}[h]
\centering
\includegraphics[width=0.50\textwidth]{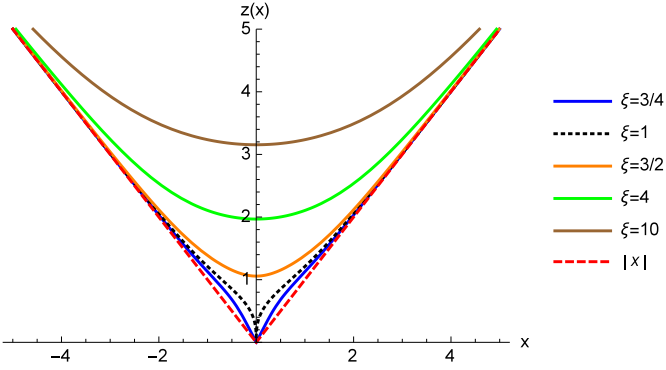}
\caption{Radial function $z(x)$ for the Case II. From bottom to top the curves represent $\xi=3/4$ (solid blue), $\xi=1$ (dotted black), $\xi=3/2$ (solid orange), $\xi=4$ (solid green) and $\xi=10$ (solid brown), with the two dashed straight red lines corresponding to $\vert x \vert $. The wormhole throat is located at $z_c=(\xi^2-1)^{1/4}$, provided that $\xi^2>1$, otherwise the radial function extends to $z=0$ (dashed black, $\xi=1$, and solid blue, $\xi=3/4$.). \label{fig:whcaseII}}
\end{figure}

Let us first study those configurations with $\xi^2>1$, for which $z_c$ has real solutions. In this case, the radial function $z(x)$ has a minimum at $z_c$ and, expanding the relation (\ref{eq:xz}) around that minimum yields
\begin{equation}
z(x) \approx z_c + \frac{z_c^4+1}{4z_c^5} x^2  \ . \label{eq:whapp2}
\end{equation}
This expansion is consistent with the bouncing behaviour depicted in full range in Fig. \ref{fig:whcaseII}, where we observe the transition between two very different behaviours for $\xi^2>1$ and $\xi^2<1$. Those with $\xi^2>1$ can be naturally interpreted as wormholes, whose radius of the throat increases as $\xi^2$ takes larger values.

To understand better the geometry at the throat $z=z_c$, we expand the relevant functions there as
\begin{eqnarray}
\Omega_{1} &\approx& \frac{2z_c^4}{z_c^4+1} -\frac{4z_c^3(z_c^4-3)}{(z_c^4+1)^2}(z-z_c)+\mathcal{O}(z-z_c)^2 \\
\Omega_{2} &\approx& \frac{4z_c^3}{z_c^4+1}(z-z_c) + \mathcal{O}(z-z_c)^2 \\
G_z&\approx& \frac{C_2}{(z-z_c)^{1/2}} + \mathcal{O}(z-z_c)^{1/2} \rightarrow \\
G(z)&\approx& -\frac{1}{\delta_c} +2C_2 (z-z_c)^{1/2} + \mathcal{O}(z-z_c)^{3/2} \ ,
\end{eqnarray}
where now the constant $C_2=\left(\frac{z_c^3}{z_c^4+1} \right)^{3/2}$ and
\begin{equation} \label{eq:deltacII}
\delta_c=-\frac{\xi ^2 \Gamma (-\frac{1}{4}) \Gamma (\frac{7}{4})}{\sqrt{2} \pi ^{3/2} z_c^3 \, _2F_1(-\frac{3}{4},\frac{1}{2};\frac{3}{4};-\frac{1}{z_c^4})} >0 \ .
\end{equation}
It should be stressed that these expressions are quite similar to those found in Eqs.(\ref{eq:Omega1a}), (\ref{eq:Omega2a}), (\ref{eq:GzcaseA}) and (\ref{eq:deltacI}) of Case I above. Moreover, the metric components take now the form
\begin{eqnarray}
g_{tt} &\approx& -\frac{r_S (\delta_1/\delta_c-1)}{4r_{\star} z_c^2 C_2\sqrt{z-z_c}} -\frac{1}{z_c C_2^{2/3}} \left(1-\frac{r_S C_2^{2/3} \delta_1}{r_{\star} z_c} \right) \\
&+& \mathcal{O}\left(\sqrt{z-z_c}\right) \nonumber \label{eq:gttCaseII} \\
g_{rr} &\approx& \frac{r_{\star}z_c^2}{r_S  C_2^{1/3} (\delta_1/\delta_c -1) \sqrt{z-z_c}} + \mathcal{O}\left(1\right) \ .
\end{eqnarray}
This is basically the same result as obtained in Eqs.(\ref{eq:gttCaseI}) and (\ref{eq:gttcaseII}) which, in turn, yields a similar structure in terms of horizons and causal regions. Moreover, curvature scalars behave in the same way, being divergent for $\delta_1 \neq \delta_c$ and finite otherwise.

\begin{figure}[h]
\centering
\includegraphics[height=5cm,width=8.5cm]{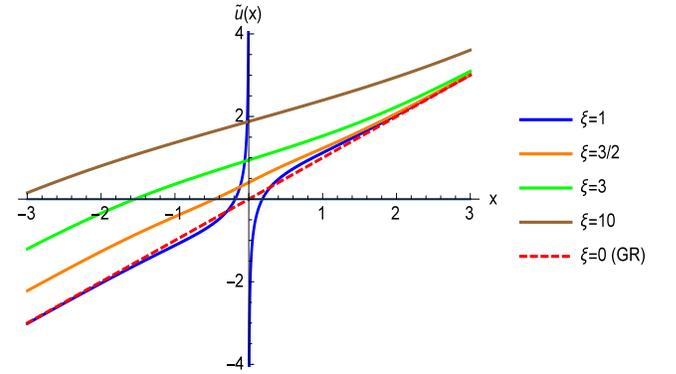}
\caption{The affine parameter $\tilde{u}(x)$ for null radial geodesics of Case II, where we verify the reliability of the approximation (\ref{eq:nullgeocaseII}) around the wormhole throat $z=z_c=(\xi^2-1)^{1/2}$ (corresponding to $x=0$ in this plot). Here we take values $\xi=3/2$ (orange), $\xi=3$ (green) and $\xi=10$ (brown), with the dashed red curve representing $\tilde{u}(x)=x$ and corresponding to the GR behaviour. These (non-GR) geodesics are complete. Moreover, we also depict the limit configuration with $\xi^2=1$, for which the wormhole throat lies at $z=x=0$, which cannot be reached in finite affine time by null radial geodesics (see Sec. \ref{sec:CaseIIc} for details). In the GR region ($z\rightarrow \infty$) all curves converge to the GR behaviour, $\tilde{u}(x) \approx x$. \label{fig:geonullII}}
\end{figure}

Regarding the behaviour of geodesics in these backgrounds, for the null ($\kappa=0$) radial ($L=0$) case we can analytically integrate the geodesic equation (\ref{eq:geoBI}) near the wormhole throat $z=z_c$, using Eqs.(\ref{eq:OmcaseIIapp}) and (\ref{eq:whapp2}), as
\begin{eqnarray}
\pm E (\tilde{u}(x)-\tilde{u}(0)) &\approx &  \frac{(z_c^4+1)^{1/2}}{z_c^{3/2}} \sqrt{z-z_c} \nonumber \\
&=&  \left(\frac{z_c^4+1}{2z_c^4}\right) x \ ,  \label{eq:nullgeocaseII}
\end{eqnarray}
which is qualitatively identical to the result obtained in Case I of section \ref{sec:CaseI}. As depicted in Fig. \ref{fig:geonullII}, where we numerically integrate the geodesic equation in all the range of definition of the radial coordinate $x$, null radial geodesics are able to cross the wormhole throat (satisfying Eq.(\ref{eq:nullgeocaseII}) there) and can be extended to arbitrarily large values of their affine parameter, thus being complete.

For null geodesics with $L \neq 0$ and time-like ($\kappa=-1$) geodesics, the fact that the expansion of the metric component $g_{tt}$ in Eq.(\ref{eq:gttCaseII}) is formally the same as that of Case I, see Eq.(\ref{eq:gttCaseI}), makes the discussion of the effective potential in the present case as equally valid as in that case. Consequently, all configurations with $\xi^2>1$ are null and time-like geodesically complete, again despite the generic existence of curvature divergences at the wormhole throat for the cases with $\delta_1 \neq \delta_c$.

\subsubsection{$\xi^2<1$} \label{sec:CaseIIb}

Let us now consider the case with $0<\xi^2<1$. Now there is no minimum $z_c$ in the radial function $z(x)$ which, consequently, runs from $(0,+\infty)$, and no wormhole is found. Expanding near the center $z=0$ one gets
\begin{eqnarray}
\Omega_{1} &\approx& \Omega_{2} \approx (1-\xi^2) + \mathcal{O}(z^4) \\
x&\approx& \sqrt{1-\xi^2}z+ \mathcal{O}(z^5) \\
G_z&\approx& \sqrt{1-\xi^2}z^2 + \mathcal{O}(z^6) \rightarrow \\
G(z)&\approx& -\frac{1}{\delta_c} + \sqrt{1-\xi^2}\frac{z^3}{3} + \mathcal{O}(z^7) \ ,
\end{eqnarray}
where the constant $\delta_c$ is now given by
\begin{equation} \label{eq:deltacIIB}
\delta_c=-\frac{\xi ^2 \Gamma (-\frac{1}{4}) \Gamma (\frac{7}{4})}{\sqrt{2} \pi ^{3/2} (1-\xi^2)^{3/4} \, _2F_1\left(-\frac{3}{4},\frac{1}{2};\frac{3}{4};-\frac{1}{1-\xi^2}\right)} >0 \ .
\end{equation}
The corresponding expansion of the metric components around the center $z=0$ yields the result (provided that $\delta_1 \neq \delta_c$)
\begin{eqnarray}
g_{tt}&\approx &-\frac{1}{1-\xi^2} + \frac{r_S(1-\delta_1/\delta_c)}{r_{\star} (1-\xi^2)^{3/2}}\frac{1}{z} +\mathcal{O}(z^2) \\
g_{rr}&\approx &\frac{r_{\star}  \sqrt{1-\xi^2}}{r_S(\delta_1/\delta_c-1)} z + \mathcal{O}(z^2) \ ,
\end{eqnarray}
and we see again that the ratio $\delta_1/\delta_c$ controls both the number of horizons and the structure of the innermost region via a similar description as in the previous case, namely, Reissner-Nordstr\"om-like configurations for $\delta_1>\delta_c$ and Schwarzschild-like solutions for $\delta_1<\delta_c$. In both cases curvature divergences of leading order $K \sim (\delta_1-\delta_c)^2/z^6$ arise at $z=0$. The absence of a wormhole implies the existence of incomplete geodesics, in much the same way as it happens in models of non-linear electrodynamics in GR.

For the case $\delta_1=\delta_c$ one must first replace this value before expanding the metric components, which yields the expressions
\begin{eqnarray}
g_{tt}&\approx& \frac{1}{1-\xi^2 }\left(-1 + \frac{r_S \delta_c}{3r_{\star}}z^2 \right) + \mathcal{O}(z^4)   \\
g_{rr}&\approx&  1+\frac{r_S \delta_c}{3r_{\star}} z^2 + \mathcal{O}(z^4) \ ,
\end{eqnarray}
and, besides finiteness of these components, one also achieves finiteness of the Kretchsman scalar, namely, $K=\frac{8r_S^2\delta_c^2}{3r_{\star}^2} + \mathcal{O}(z^2)$. In addition, it can be verified that the geometry in this case around $z=0$ satisfies
\begin{equation}
R_{\mu\nu}=\Lambda_{eff} g_{\mu\nu} \ ,
\end{equation}
which is of de Sitter type, with effective cosmological constant $\Lambda_{eff}=\frac{r_S\delta_c}{r_{\star}^3}$\footnote{It should be pointed out that the development of a de Sitter core has been known for quite some time ago to be a mechanism able to get rid of curvature divergences, which has shaped many approaches to this issue in the context of GR \cite{dS}.}. In addition, a simple re-scaling of the time coordinate of the form $t \rightarrow \sqrt{1-\xi^2} t$ brings the corresponding line element into a locally Minkowskian form.

The de Sitter core puts forward that the geometry is smooth in the central region. In fact, null radial geodesics around  $z=0$ behave as
\begin{equation}
\pm E \frac{d\tilde{u}(x)}{dz} \approx \frac{1}{\sqrt{1-\xi^2}} + \mathcal{O}(z^4) \ ,
\end{equation}
whose integration yields $\pm E (\tilde{u}(x)-\tilde{u}(0))= \frac{z}{\sqrt{1-\xi^2}}$ (the $\pm$ sign denotes outgoing/ingoing trajectories). This result implies that an ingoing ray can reach $z=0$ in a finite affine time. At that point, the ingoing ray turns into outgoing, flipping the sign of $(\tilde{u}(x)-\tilde{u}(0))$ and allowing for its extension to arbitrarily large values, thus confirming the completeness of these geodesics. Given the timelike character of the surface $z=0$, similar conclusions follow for the other geodesics (non-radial and timelike).

\subsubsection{$\xi^2=1$} \label{sec:CaseIIc}

Let us finally analyze the limiting case $\xi^2=1$. Now, the expansion of the relevant functions around $z=0$ yields
\begin{eqnarray}
\Omega_{1} &\approx& 3z^4-5z^8 + \mathcal{O}(z^{12}) \\
\Omega_{2} &\approx& z^4-z^8 +  \mathcal{O}(z^{12}) \\
G_z&\approx& 3z^4-\frac{13}{2} z^8 +  \mathcal{O}(z^{12}) \\
G(z)&\approx& -\frac{1}{\delta_c} + \frac{3}{5}z^5 +  \mathcal{O}(z^{9}) \\
g_{tt}&\approx& \frac{r_S (1-\delta_1/\delta_c)}{3r_{\star}z^7} -\frac{1}{3z^4} +  \mathcal{O}(z^{-3}) \\
g_{rr}&\approx& \frac{3r_{\star}}{r_S(\delta_1/\delta_c-1)} z^3 + \mathcal{O}(z^6) \ ,
\end{eqnarray}
where now the constant $\delta_c=3\Gamma[3/4]^2/\pi^{3/2} \simeq 0.80902$. Should we try in this case to sustain the wormhole interpretation of the cases with $\xi^2>1$, then the expansion of the radial function would yield
\begin{equation}
x \approx z^3 - \frac{z^7}{2} +  \mathcal{O}(z^{-11}) \ .
\end{equation}
This implies that the wormhole throat in this case would have vanishing area, see Fig. \ref{fig:whcaseII} (dashed black curve). Moreover, curvature divergences always arise at $z=0$, being of order $\sim (\delta_1-\delta_c)^2/z^{10}$ in general and softened to $\sim 1/z^4$ when $\delta_1=\delta_c$.

Regarding geodesic behaviour, for any $\delta_1$, null radial geodesics integrate the equation  (\ref{eq:geoBI}) near the center $z\approx x \approx 0$ as
\begin{equation}
\pm E (\tilde{u}(x)-\tilde{u}(0))\approx  - \frac{1}{z} + \mathcal{O}(z^3) \ .
\end{equation}
As depicted in Fig. \ref{fig:geonullII} (solid blue), this result implies that the throat cannot be reached in finite affine time by null radial geodesics. This result is similar to that found in the case of certain Palatini $f(R)$ theories coupled to electromagnetic fields \cite{ORUniverse} or to anisotropic fluids \cite{BOR2017}.

The analysis of the effective potential around $z\approx x \approx 0$ in this case, $V_{eff} \approx - \frac{\tilde{a}}{\vert x \vert^7} - \frac{\tilde{b}}{\vert x \vert^4 }$  (with $\tilde{a}=\frac{r_S}{3r_{\star}} (1-\delta_1/\delta_2)(L^2/x^2-\kappa)$ and $\tilde{b}=-1/3(L^2/x^2-\kappa)$)  for null (with $L\neq 0$) and timelike geodesics reveals a similar fate for them as for those of Case I: those with $\delta_1>\delta_c$ (Reissner-Nordstr\"om-like configurations) will find an infinitely repulsive potential barrier and be scattered off to asymptotic infinity, while those with $\delta_1<\delta_c$ (Schwarzschild-like configurations) will be dragged towards the wormhole throat $x=0$. In the latter case, a curious effect occurs, since the integration of the geodesic equation yields the result that the region $x=0$ can be reached in finite affine time both for null (with angular momentum) and timelike geodesics (for example, timelike radial geodesics behave there as $\pm (\tilde{u}(x)- \tilde{u}(0) )\approx \xi \vert x \vert^{5/6}$, with $\xi$ some constant), despite the infinite time required by null radial geodesics to get there. The vanishing area of the wormhole suggests that extended objects would be compressed to zero volume as the throat is reached, indicating that such solutions are pathological.

\subsection{Case III: $\{s_{\epsilon}=+1,s_{\beta}=-1\}$} \label{sec:CaseIII}

\begin{figure}[h]
\centering
\includegraphics[height=6cm,width=8.5cm]{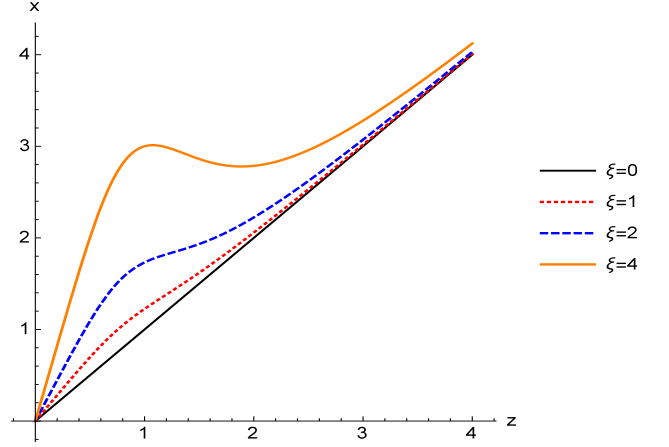}
\caption{Representation of $x(z)$ of Case III (corresponding to $\{s_{\epsilon}=+1,s_{\beta}=-1\}$) for $\xi=0$ (solid black), $\xi=1$ (dotted red), $\xi=2$ (dashed blue) and $\xi=4$ (solid orange). \label{fig:xz0}}
\end{figure}

Now we have the expressions
\begin{eqnarray}
\rho&=&\frac{\rho_m}{z^4+1} \\
\Omega_{1}&=&1-\frac{\xi^2(z^4-1)}{(z^4+1)^2} \hspace{0.1cm};\hspace{0.1cm} \Omega_2=1+\frac{\xi^2}{z^4+1} \\
G_z&=&\frac{z^2 \Omega_1}{(z^4+1) \Omega_{2}^{1/2}} \ .
\end{eqnarray}
In this case, inspecting the relation (\ref{eq:xz}) it turns out that, for $\xi^2 \geq 8$, the function $x(z)$ has a minimum at $\gamma=\frac{(z_{min}^4+1)^2}{z_{min}^4-1}$, in such a way that the function $G_z$ has two zeros, located at $0<z_{max} \leq 1 \leq z_{min}$, corresponding to a local maximum and minimum, respectively (see Fig. \ref{fig:xz0}). However, the presence of such a minimum in $x(z)$ cannot be interpreted as representing a wormhole throat in the auxiliary metric $q_{\mu\nu}$, since it does not correspond to an absolute minimum. Indeed, the function $x(z)$ can be extended in a monotonic way below $x_{max}$ all the way down to $x=0$. In that region the relation (\ref{eq:xz}) becomes
\begin{equation}
x \approx \sqrt{1+\xi^2}z\left(1-\frac{\xi^2}{(1+\xi^2)}z^4 + \mathcal{O}(z^8)   \right) \ ,
\end{equation}
so $z(x) \simeq x/(\sqrt{1+\xi^2})$ there, which amounts just to a re-scaling of the radial coordinate. This result is consistent with the numerical integration depicted in Fig. \ref{fig:xz0}, which does not correspond to the expected smooth bouncing behaviour of a wormhole structure, as follows from the fact that no zeros can be found for $\Omega_2$ in this case.

The lack of a wormhole structure suggests that one should focus on the parametrization of the solutions in terms of $z$.
To further understand  the geometry at $z=0$ we follow a similar strategy as in the previous sections, where we expand the relevant functions there, which in this case yields the metric components
\begin{eqnarray}
g_{tt}&\approx &-\frac{1}{1+\xi^2} + \frac{r_S(1-\delta_1/\delta_c)}{r_{\star} (1+\xi^2)^{3/2}}\frac{1}{z} +\mathcal{O}(z^2) \\
g_{rr}&\approx &-\frac{r_{\star}\sqrt{1+\xi^2}}{r_S(1-\delta_1/\delta_c)} z + \mathcal{O}(z^2) \ ,
\end{eqnarray}
which is the same result as in case $0<\xi^2<1$ of section \ref{sec:CaseII} with the replacement $\xi^2 \rightarrow -\xi^2$. Therefore, similar comments regarding the features of the corresponding solutions apply (such as the number and type of horizons) and, in particular, solutions with $\delta_1=\delta_c$ represent de Sitter cores at the center, $R_{\mu\nu}=\frac{r_S \delta_c}{r_{\star}^3} g_{\mu\nu}$,  with all curvature scalars being finite (while divergences of order $\sim (\delta_1-\delta_c)^2/z^6$ arise for $\delta_1 \neq \delta_c$).

In this case the integration of the geodesic equation (\ref{eq:geoBI}) around $z=0$ in the null radial case yields the result $\pm E (\tilde{u}(x)-\tilde{u}(0)) \approx  \frac{z}{\sqrt{\xi^2+1}}$, which implies that $z=0$ can be reached in finite affine time. As a result, in the Schwarzschild-like solutions  there is no possibility of extending those geodesics, implying that these geometries are geodesically incomplete.

\subsection{Case IV: $\{s_{\epsilon}=+1,s_{\beta}=+1\}$} \label{sec:CaseIV}

\begin{figure}[h]
\centering
\includegraphics[width=0.48\textwidth]{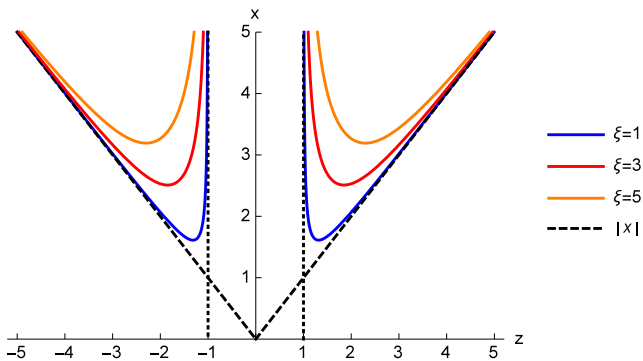}
\caption{Representation of $x(z)$ of Case IV for $\xi=1$ (solid blue), $\xi=3$ (solid red), $\xi=5$ (solid orange), as compared to $\vert x \vert=z$ (dashed black). The vertical dotted lines set the minimum radius available for the radial function, $z=1$, where the energy density diverges.  \label{fig:xz}}
\end{figure}

In this case we have the expressions
\begin{eqnarray}
\rho&=&\frac{\rho_m}{z^4-1} \label{eq:rhoIV}\\
\Omega_{1}&=&1-\frac{\xi^2(z^4+1)}{(z^4-1)^2} \hspace{0.1cm};\hspace{0.1cm} \Omega_{2}=1+\frac{\xi^2}{z^4-1}\\
G_z&=&\frac{z^2 \Omega_{1}}{(z^4-1)\Omega_{2}^{1/2}}
\end{eqnarray}
Here we see some differences as compared to the previous cases: for $\xi^2 \leq 1$ one finds that $\Omega_2$ vanishes at a value $z_c=(1-\xi^2)^{1/4}<1$, which lies beyond the point at which the energy density of the fluid blows up, $z=1$ (note that $\Omega_1$, $\Omega_2$ and $G_z$ blow up there too), while for $\xi^2 >1$ no real value for $z_c$ can be found. In all cases, at the limiting  radius $z=1$ one can see that the relation (\ref{eq:xz}) between radial coordinates blows up, leading to the behaviour depicted in Fig. \ref{fig:xz}.

Exploring further the nature of the surface $z=1$, we expand the metric functions there to find
\begin{eqnarray}
G(z)&\approx& \frac{\xi}{24(z-1)^{5/2}} + \mathcal{O}(z-1)^{-3/2} \\
g_{tt}&\approx& -\frac{2r_S\delta_1}{3r_{\star}\xi^2} (z-1) + \mathcal{O}(z-1)^2 \\
g_{rr}&\approx& \frac{6r_{\star}}{r_S \delta_1} + \mathcal{O}(z-1)^2 \ .
\end{eqnarray}
Despite the finiteness of $g_{tt}$ and $g_{rr}$ at the surface $z=1$, curvature divergences with strength $\sim 1/(z-1)^4$ arise there, which cannot be avoided for any choice of $\delta_1$, unlike the Cases I and II above. When null radial geodesics are considered, the geodesic equations can be integrated near the surface $z=1$ as
\begin{eqnarray}
\pm E (\tilde{u}(x)-\tilde{u}(0)) \approx \frac{4(z-1)^{3/2}}{3\xi} + \mathcal{O}(z-1)^{5/2} \ ,
\end{eqnarray}
which implies that these geodesics can reach $z=1$ in finite affine time. Although, in principle, it should be possible to construct an analytical extension to the inner region $z<1$, the fact that the energy density of the fluid blows up at $z=1$ suggests the breakdown of the matter description there and, consequently, the non-physical character of such an extension. We shall thus leave it here, and just mention that an extension of this kind was recently constructed in the context of Palatini $f(R)$ gravity \cite{BOR2017}.

\section{Conclusions} \label{sec:VI}

In this work we have considered an extension of GR known as Eddington-inspired Born-Infeld gravity coupled to an anisotropic fluid constrained to satisfy reasonable physical conditions and incorporating a number of interesting scenarios, such as those of nonlinear electrodynamics. Focusing on static, spherically symmetric solutions, we have solved the field equations in closed form. The combination of the signs of two parameters in the gravity sector, $\epsilon$, and the fluid description, $\beta$, led us to split the analysis into four cases, where a variety of different configurations are found. All of them recover the Reissner-Nordstr\"om solution of the Einstein-Maxwell field equations at large distances, but important departures with respect to that solution are found as we approach the innermost region. On each branch, the modifications with respect to GR due to the interplay between gravity and matter are encoded on a single scale, $\xi^2$.

The most physically appealing results are found in the $\epsilon<0$ branch. For both $\beta>0$ (Case I) and $\beta<0$ (Case II) the energy density of the fluid is finite everywhere and the point-like GR singularity is replaced by a wormhole structure with its throat located at $z=z_c=(1+\xi^2)^{1/4}$ (in Case I), and at $z=z_c=(\xi^2-1)^{1/2}$ (in Case II, provided that $\xi^2>1$). In these cases, the ratio $\delta_1/\delta_c$, where $\delta_1$ encodes the relevant parameters of the solutions and $\delta_c$ is a constant, plays a key role in the number and type and horizons and, consequently, on the causal structure of the solutions. Indeed, if $\delta_1>\delta_c$ the configurations have the typical structure of the Reissner-Nordstr\"om solution of GR, namely, either black holes with two horizons, extreme black holes (a single degenerate horizon) or no horizons, while for $\delta_1<\delta_c$ a Schwarzschild-like configuration is found, always characterized by the presence of a single (non-degenerate) horizon. Finally, those configurations with $\delta_1=\delta_c$ may have an event horizon or none, but the metric component $g_{tt}$ is always finite at the center of the solutions. This description of horizons mimics that found in certain models of nonlinear electrodynamics in the context of GR (see \cite{NEDg-GR} for a detailed discussion on that issue).

The presence of a wormhole structure in all the configurations of Case I and in those of Case II with $\xi^2>1$ has also a non-negligible impact on the regularity of the solutions. Indeed, in Case I the strength of the curvature divergences at the wormhole throat softens from the $\sim 1/z^8$ behaviour of the (charged) black hole of the GR case, to a leading-order divergence $\sim (\delta_1-\delta_c)/(z-z_c)^3$. Moreover, due to the dependence on the ratio $\delta_1/\delta_c$, it also follows that when $\delta_1=\delta_c$ then we get rid of any divergences on curvature scalars. To delve deeper into the implications of this result we have studied the geodesic motion on these geometries, making use of the standard approach for this problem suitably adapted to Palatini theories of gravity. This way, we have found that radial null geodesics are able to reach the wormhole throat in finite affine time but they are naturally extended beyond that point, which contrast with the termination of the geodesics there in the GR case. In addition, we have considered null (with $L \neq 0$) and timelike geodesics, formulating the problem in a way akin to the motion of a one-dimensional particle in an effective potential. In the $\delta_1>\delta_c$ this potential prevents any such geodesic to reach the wormhole throat, in much the same way as in the Reissner-Nordstr\"om case of GR. However, for $\delta_1 \leq \delta_c$ these geodesics may reach the wormhole throat in finite affine time (depending on their energy $E$) but, like the radial null ones, can be smoothly extended beyond that point. Therefore these solutions are null and timelike geodesically complete. The regularity of many of the solution of this $\epsilon <0$ branch is consistent with previous analyses of geodesic completeness of this branch in the context of electrovacuum solutions with Maxwell fields \cite{ORS16a,ORS16b}.

For Case II with $\xi^2<1$ there is no wormhole and the radial function extends all the way down to $z=0$, which can be reached in finite affine time by null radial geodesics with no possibility of further extension. Hence, these solutions are geodesically incomplete in general. As an exception to the general case, we find that a de Sitter core arises when $\delta_1 = \delta_c$, which regularizes all curvature scalars and guarantees the extendibility of geodesics.
More striking results are found when $\xi^2=1$, since in such a case null radial geodesics take an infinite affine time to get to $z=0$ (this result being very similar to those found in some $f(R)$ models in Palatini formulation \cite{ORUniverse,BOR2017}), but null (with $L \neq 0$) and timelike geodesics in the Schwarzschild-like configurations ($\delta_1<\delta_c$) may get there in finite affine time, though its extendibility beyond that point is unclear due to the vanishing area of the wormhole throat in this case.

For $\epsilon>0$ and $\beta<0$ (Case III) the energy density is finite, and no wormhole structure is found, though we still have the description of horizons parameterized by the ratio $\delta_1/\delta_c$. Due to the lack of a wormhole, the fact that $z=0$ can be reached in finite affine time by null radial geodesics implies the incompleteness of geodesics. Finally, for $\epsilon>0$ and $\beta>0$ (Case IV), the energy density of the fluid blows up at the surface $z=1$, where curvature divergences arise and which is reached in finite affine time by null radial geodesics. In this case, the breakdown in the description of the fluid suggests the non-physical character of the $z<1$ region, despite the fact that, in principle, analytical extensions of the metric to this region could be possible.

In summary, we have found several physically appealing structures that include nonsingular black holes and nonsingular naked compact objects, wormholes and de Sitter cores. Such objects are the result of the non-trivial interaction between gravity and matter ascribed to Palatini theories of gravity, where the energy density of the matter fields introduces additional effects on how these fields gravitate as compared to GR. Our findings are added to the growing set of results within this kind of theories, where the GR point-like singularity may be replaced by a geodesically complete spacetime using matter sources that satisfy the standard classical energy conditions. Moreover, these geometries break the correlation between geodesic completeness and curvature divergences, since the latter do not prevent the former. In this sense, we note that in the context of EiBI gravity with electromagnetic fields, it has been found that extended bodies and waves crossing the wormhole throat do not experience any kind of pathological or destructive effects \cite{ors16}. Further research on the behaviour of classical and quantum fields, as well as the propagation of gravitational waves on these backgrounds is necessary, and we hope to report on these topics soon.

\section*{Acknowledgments}

C. M. and G. J. O. are funded by the fellowship No.~BES-2015-072941 and a Ramon y Cajal contract, respectively. D. R.-G. is funded by the Funda\c{c}\~ao para a Ci\^encia e a Tecnologia (FCT, Portugal) postdoctoral fellowship No.~SFRH/BPD/102958/2014 and the FCT research grant UID/FIS/04434/2013. This work is supported by the Spanish grant FIS2014-57387-C3-1-P (MINECO/FEDER, EU), the project H2020-MSCA-RISE-2017 Grant FunFiCO-777740, the project SEJI/2017/042 (Generalitat Valenciana), the Consolider Program CPANPHY-1205388, and the Severo Ochoa grant SEV-2014-0398 (Spain). This article is based upon work from COST Action CA15117, supported by COST (European Cooperation in Science and Technology).

\end{document}